\documentclass[twocolumn,showpacs,preprintnumbers,amsmath,amssymb]{revtex4}
\usepackage{graphicx}% Include figure files
\usepackage{dcolumn}% Align table columns on decimal point
\usepackage{bm}% bold math

\begin{document}
\title{ Transport anomalies across the quantum limit in semimetallic Bi$_{0.96}$Sb$_{0.04}$}
\author{Aritra Banerjee$^{1}$, Beno\^{\i}t Fauqu\'e$^{1}$, Koichi Izawa$^{2,3}$ Atsushi Miyake$^{2}$,
Ilya Sheikin$^{4}$, Jacques Flouquet$^{2}$, Bertrand Lenoir$^{5}$
and Kamran Behnia$^{1}$} \affiliation{(1)LPEM (UPMC-CNRS), Ecole
Sup\'erieure
de Physique et de Chimie Industrielles, 75231 Paris, France \\
(2)DRFMC/SPSMS,  Commissariat \`a l'Energie Atomique, 38042
Grenoble, France\\
(3)Graduate School of Science and Engineering, Tokyo Institute of
Technology, Tokyo 152-8551, Japan\\
(4) Grenoble High Magnetic Field Laboratory (CNRS), BP 166, 38042
Grenoble , France\\
(5)Laboratoire de Physique des Mat\'eriaux  (Nancy Universit\'e,
CNRS, Ecole des Mines de Nancy), Parc de Saurupt, 54042 Nancy,
France }

\date {July 24, 2008}

\begin{abstract}
We report on a study of electronic transport in semi-metallic
Bi$_{0.96}$Sb$_{0.04}$. At zero field, the system is a very dilute
Fermi liquid displaying a T$^{2}$ resistivity with an enhanced
prefactor. Quantum oscillations in resistivity as well as in Hall,
Nernst and Seebeck responses of the system are detectable and their
period quantifies the shrinking of the Fermi surface with antimony
doping. For a field along the trigonal axis, the quantum limit was
found to occur at a field as low as 3T. An ultraquantum anomaly at
twice this field was detected in both charge transport and Nernst
response. Its origin appears to lie beyond the one-particle picture
and linked to unidentified many-body effects.

\end{abstract}

\pacs{71.70.Di, 71.18.+y, 72.15.Jf, }

\maketitle

It has been known for a long time that doping elemental bismuth with
antimony decreases the carrier density and eventually leads to a
semimetal to semiconductor transition\cite{jain,lerner}.
Bi$_{x}$Sb$_{1-x}$ alloys are remarkable n-type thermoelectric
material operating near 100 K\cite{smith1}. The evolution of their
Fermi surface \cite{brandt1}, notably under
pressure\cite{brandt2,brandt3}, as well as their transport
coefficients above 4.2 K\cite{lenoir} have been intensively studied.
They are attracting new attention in the context of research on
``topological insulator'', which was first proposed \cite{fu} and
then reported to be detected\cite{hsieh} in this family on the
semiconductor side (i.e. for $x>0.08$). The metallic side of this
semimetallic-semiconductor phase boundary permits to explore the
instabilities of an ambipolar three-dimensional electron gas as the
carrier density is continuously pushed to zero\cite{halperin1}.
Moreover, the small size of the Fermi surface pulls down the quantum
limit. This limit is attained when all carriers are at the lowest
Landau level. The barely-screened Coulomb repulsion of the semimetal
at zero-field is expected to become even stronger beyond this limit.
In the case of pure bismuth,  the quantum limit is attained by a
field of 9T applied along the trigonal axis\cite{bompadre,behnia1}.
The fate of the three-dimensional electron gas beyond this field,
invoked years ago from a theoretical point of view\cite{halperin2}
has been barely explored experimentally in spite of a number of
high-field studies\cite{hiruma1,hiruma2}. A recent study of pure
bismuth detected unexpected transport anomalies in this
regime\cite{behnia2} indicating unknown many-body effects. The
decrease in carrier density induced by Sb doping should push the
quantum limit to even lower and more accessible fields. A last
motivation is provided by the Dirac fermions, which are expected to
become massless with the closing of the L-point gap at
x=0.04\cite{mikitik,ghosal}.
\begin{figure}
 \resizebox{!}{0.62\textwidth}{\includegraphics{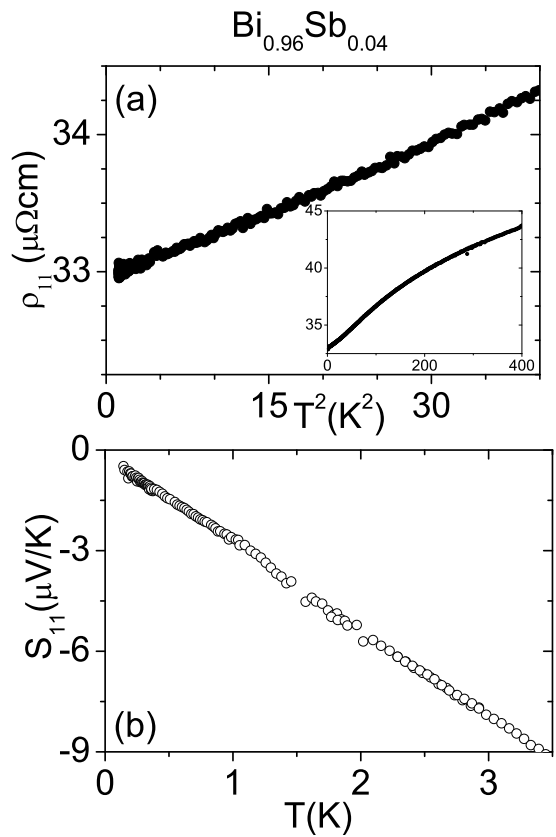}}
\caption{\label{Fig1} (a) The zero-field resistivity
Bi$_{0.96}$Sb$_{0.04}$ as a function of T$^{2}$ for a current along
the binary axis. The inset shows the data over a wider temperature
range. (b)The temperature-dependence of the Seebeck coefficient at
low temperatures for a temperature gradient along the binary axis.}
\end{figure}
In this Letter, we present a study of electric and thermoelectric
transport in semimetallic Bi$_{0.96}$Sb$_{0.04}$ extended to low
temperatures (0.15 K) and high magnetic field (28 T). In zero
magnetic field, we find two hallmarks of a Fermi-liquid behavior
namely, a T$^2$ resistivity and a T-linear Seebeck coefficient. The
magnitude of both point to a remarkably small Fermi temperature. The
electronic mobility, while significantly reduced compared to pure
bismuth, remains high in this alloy. Then we present quantum
oscillations of transport properties (and in particular the Nernst
coefficient), whose period quantifies the size of the Fermi surface.
The quantum limit, for a field along trigonal occurs at 3 T, i.e.
three times lower than in pure bismuth. Finally, we resolve
anomalies in all transport properties occurring at a field of 6 T.
This field scale, which cannot be attributed to the crossing of the
Fermi level by a known Landau level is reminiscent of (and different
in position with) the case of pure bismuth \cite{behnia2}. Its
explanation appears to be beyond the one-particle picture.

Single crystals of Bi$_{1-x}$Sb$_{x}$  were prepared using a
traveling heater method as detailed in ref. \cite{lenoir}. The
nominal concentration of antimony in the single crystalline samples
studied in this work was 4 percent.  Assuming that the lattice
parameter is a linear function of $x$ and by measuring it in a small
piece cut from the sample by powder diffractometry, we found an
antimony concentration of $x=0.037$. The longitudinal electrical and
thermal conductivity, as well as Hall, Seebeck and Nernst
coefficients of the system were measured using a standard 6-wire
method with two thermometers and one heater in a 12 T
superconducting magnet for two different orientations of  magnetic
field. The results were complemented by a set of measurements for
the trigonal configuration with a 28 T resistive magnet.

Fig. 1 summarizes the transport properties of the system in the
zero-field limit. As seen in panel $a$, resistivity is a quadratic
function of temperature ($\rho=\rho_{0}+AT^{2}$), the expected
behavior of a Landau Fermi liquid. The magnitude of the $A$
coefficient ($33 n\Omega cm K^{-2}$ in the $T<5K$ range ) is almost
three times larger than in pure bismuth($12 n\Omega cm
K^{-2}$)\cite{hartman,uher}. The enhanced $A$ is a natural
consequence of the reduction in the Fermi temperature. It is
instructive to contrast the magnitude of $A$ (remarkably large for
these light-electron semimetals) with the modest magnitude of the
electronic specific heat, which has been reported to be as low as
$\gamma=8.5 \mu J mol^{-1}K^{-1}$in pure bismuth\cite{collan}. Thus,
the Kadowaki-Woods (KW) ratio
($\frac{A}{\gamma^{2}}$\cite{kadowaki}) in semimetallic bismuth is
five orders of magnitude larger than in high-carrier density metals
(i.e. those with roughly one itinerant electron per formula unit).
Theoretical examinations of the KW ratio had already concluded that
this ratio should increase as the Fermi surface decreases in
size\cite{kontani,hussey}. Its exceptionally large value in bismuth,
however, had hitherto escaped attention.

As seen in the panel b of the same figure, the Seebeck coefficient
is negative and quasi-linear in temperature below 3 K. Our data
smoothly join the reported high-temperature behavior\cite{lenoir}.
In contrast with pure bismuth, no remarkable structure at low
temperature emerges and a large slope ($\frac{S}{T}\simeq-2.7\mu
VK^{-2}$), the expected consequence of the small Fermi temperature
can be easily extracted. We note that the dimensionless ratio of
linear thermopower to linear specific heat ($q=\frac{S}{T}\frac{e
N_{Av}}{\gamma}$), where $e$ is the electron charge and $N_{Av}$ the
Avogadro number) is more than four orders of magnitude larger than
unity, the typical ratio found in metals with conventional carrier
density\cite{behnia3}. Since $q$ is expected to inversely scale with
carrier density\cite{behnia3}, this is not surprising. It is
instructive to compare the low-temperature thermoelectric responses
of Bi$_{0.96}$Sb$_{0.04}$ and Bi. In pure bismuth\cite{uher2}, it is
non-monotonous and very small below 1K indicating that contributions
by electrons  and holes to thermopower are comparable and cancel out
at very low temperatures. On the other hand, the frankly negative
and $T$-linear Seebeck coefficient in Bi$_{0.96}$Sb$_{0.04}$ implies
the domination of electron-like carriers. Since the system remains
compensated the contraction of the volume of the electron and hole
pockets should be strictly equal. Therefore, this predominance
cannot be simply explained by the difference in the evolution of the
hole and electron Fermi energies with Sb doping.  On the other hand,
close to x=0.04, the energy spectrum of the electron pocket, which
is only partly linear in momentum in the case of pure bismuth
becomes strictly linear. Thus, the negative sign of the thermopower
reflects the enhanced thermoelectric response of Dirac fermions
compared to conventional quasi-particles.

We also measured the thermal conductivity of the sample in zero field (not
shown) and found that it is dominated by lattice thermal
conductivity as in pure bismuth with an almost undetectable electron
contribution\cite{behnia4}.

\begin{figure}
 {\includegraphics{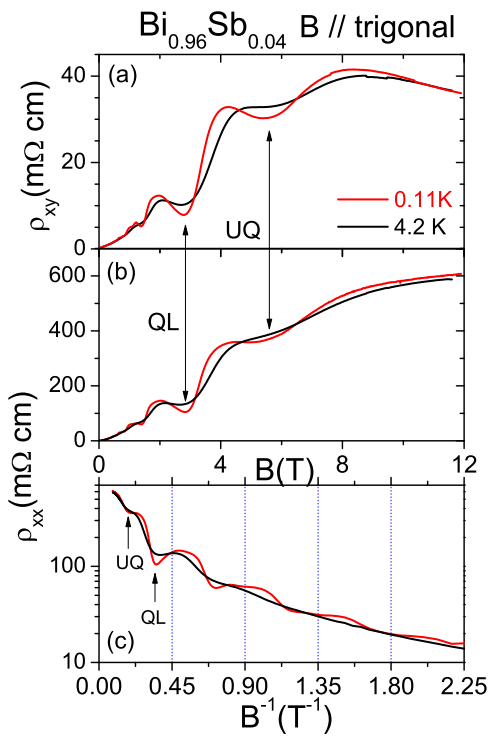}}
\caption{\label{Fig2} The Hall(a) and the longitudinal resistivity
(b) as a function of magnetic field at two different temperatures.
QL marks the quantum limit anomaly. Beyond this limit (i.e. in the
ultraquantum regime), another anomaly is visible and marked as UQ.
(c) presents a semi-log plot of resistivity as a function of
B$^{-1}$. Quantum oscillations are visible. }
\end{figure}
To probe the Fermi surface quantitatively, we studied the quantum
oscillations of various transport properties. As seen in Fig.2, in
presence of a quantizing magnetic field along the trigonal, both
resistivity and the Hall effect show visible oscillations. Panel c
of the same figure, a semi logarithmic plot of resistivity $vs.$
inverse of the magnetic field, reveals an oscillation  period of
0.45 $T^{-1}$, three times larger than in pure bismuth
(0.15T$^{-1})$\cite{bompadre,behnia1}. This means that the
equatorial cross section of the hole ellipsoid has become three
times smaller. The quantum limit, the passage to the $n=0$ Landau
level, occurs at field of about 3T. Note also the additional anomaly
in both $\rho_{xx}$ and $\rho_{xy}$ at a yet higher field ($\sim 6
T$).

\begin{figure}
 {\includegraphics{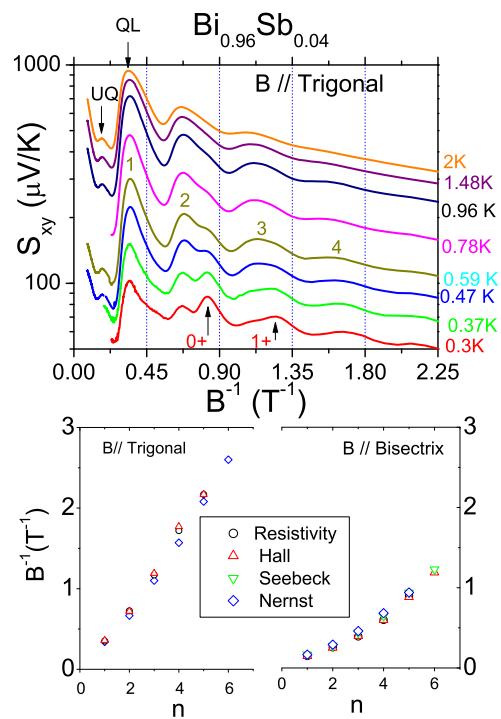}}
\caption{\label{Fig3} Upper Panel: Nernst signal as a function of
inverse of the magnetic field.  Numbers identify peaks associated
with various Landau levels. Note the gradual emergence of new peaks
due to Zeeman splitting at lower temperatures (See text). The lower panels present the
position of identified landau levels for two different field orientations. }
\end{figure}

As seen in the main panel of Fig. 3, quantum oscillations of the
Nernst response are easily detectable. Their large amplitude is
reminiscent of (but less pronounced than) the case of pure
bismuth\cite{behnia1}. The period of these oscillations matches the
period of Shubnikov-de Haas oscillations of Fig.2. Each Nernst
maximum is concomitant with a resistivity minimum. One remarkable
feature is the emergence of new peaks at the lowest temperature
linked to the large Zeeman splitting. In bismuth for a field along
trigonal, the Zeeman energy, E$_{Z}$, of holes is such that
E$_{Z}=2.17\hbar\omega_{c}$ \cite{bompadre}. In other words, the
Zeeman splitting is so large that the field corresponding to the
$0^{+}$ peak is $lower$ than the one associated with $2^{-}$. This
was detected in both resistivity\cite{bompadre} and
Nernst\cite{behnia1} data. Thus, the occurrence of the same feature
in our data on Bi$_{0.96}$Sb$_{0.04}$ indicates that here also
E$_{Z} > 2 \hbar\omega_{c}$. As the system cools, the $0^{+}$
anomaly gradually dominates the $2^{-}$ peak. We note that this
temperature evolution is more radical here than in the case of pure
bismuth \cite{behnia1}.

The period of quantum oscillations varies as a function of magnetic field. This can
be seen in the lower panels of Fig.3 which presents the B$^{-1}$ position of successive
Landau levels for both orientations of magnetic field.
If the period was constant, the data would fall on a straight line. The upward curvature
is reminiscent of the case of pure bismuth\cite{bompadre} and is a signature of a field-induced
modification of the carrier density and Fermi surface volume. According to a
picture proposed many years ago\cite{smith2}, charge neutrality in
this compensated system implies a continuous adjustment in carrier density of both holes
and electrons. This leads to a steady increase in the size of the Fermi surface, in particular
in the vicinity of the quantum limit. As seen in the lower panels, the upward curvature is
much stronger for a field along the bisectrix, which was also the case in pure bismuth\cite{smith2}.

\begin{table}

\begin{tabular}{|c|c|c|c|}
  \hline
System & $P_{3}^{h}(T^{-1})$ & $P_{1}^{h}(T^{-1}$)  & FS Vol. (T$^{3/2}$) \\

\hline
Bi  & 0.157 &   0.045 & 30.0 \\
\hline
Bi$_{0.96}$Sb$_{0.04}$  &0.5 & 0.3& 3.64  \\
\hline

\end{tabular}
\caption{Period of the quantum oscillations for a field along the
trigonal axis, $P_{3}^{h}$, and perpendicular to it, $P_{1}^{h}$,
together with the calculated size of the Fermi surface in
Bi\cite{bhargava} and in Bi$_{0.96}$Sb$_{0.04}$. The values
correspond to the low-field limit. The Fermi surface volume is
calculated as $\frac{1}{P_{3}^{h}\sqrt{P_{1}^{h}}}$.}\label{T1}
\end{table}

 Table 1 compares the period of quantum oscillations in Bi (according to one authoritative
studiy\cite{bhargava}) with what is found here for
Bi$_{0.96}$Sb$_{0.04}$ at the low field. For each field orientation,
the frequency of oscillations directly yields the projected area of
the Fermi surface. Hence, the volume of the hole ellipsoid can be
unambiguously determined. It is smaller by a factor of 8 in the
alloy. Even though the electron pockets have remained invisible,
their overall volume is expected to contract identically. Indeed,
the system is expected to remain compensated during the smooth
evolution towards the insulator. Thus, we conclude that the carrier
density for both electrons and holes in Bi$_{0.96}$Sb$_{0.04}$ is
$4\times 10^{16}cm^{-3}$. This is in fair agreement with the
estimation by Brandt and co-workers\cite{brandt3}. It is remarkable
to find that a Fermi liquid behavior persists at such a low level of
carrier concentration. A residual resistivity of 32 $\mu\Omega cm$
for such carrier density (assuming an equal mobility for electrons
and holes) would imply a mobility of $2.4 \times 10^{6}
cm^{2}V^{-1}s^{-1}$, significantly lower than bismuth\cite{hartman},
but remarkably high for an alloy .

\begin{figure}
 {\includegraphics{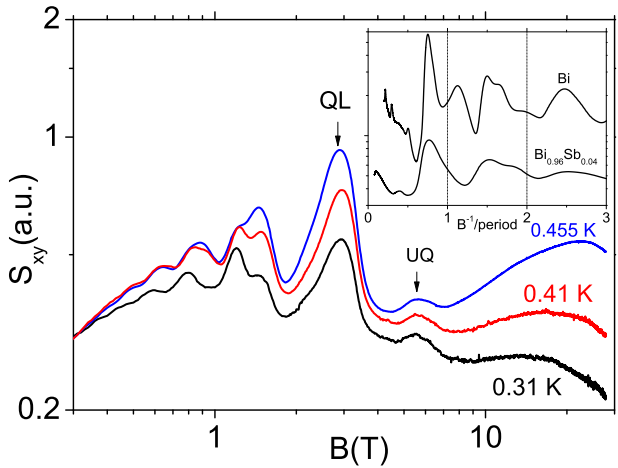}}
\caption{\label{Fig4} Nernst data obtained for
Bi$_{0.96}$Sb$_{0.04}$ in a high-field (28T) resistive magnet.  The
insert compares the alloy with pure bismuth. The  0.45K data  and
the 0.85K data for Bi\cite{behnia2} is plotted  as a function of $
B^{-1}$ normalized to the period of oscillations in each case (See
text). }
\end{figure}

In order to investigate the transport properties of the system deep
in the ultraquantum regime, the Bi$_{0.96}$Sb$_{0.04}$ crystal was
put in a 28 T resistive magnet of Grenoble High Magnetic Field Lab
with a set-up designed for high-resolution thermoelectric
measurements in a top-loading dilution refrigerator. The results are
presented in Fig.4. As seen in  the figure, in addition to the
ultraquantum anomaly at 6 T, the data presents a very broad maximum
at $\sim$ 18 T. In contrast to the 6T anomaly, the field position of
the latter feature evolves with cooling and one cannot firmly
identify a zero-temperature field scale.

The inset of the figure compares the Nernst anomalies in bismuth\cite{behnia2} and in
Bi$_{0.96}$Sb$_{0.04}$. The horizontal axis is  $B^{-1}$ normalized by the oscillation
period in each system (0.15 T$^{-1}$ in the pure case and 0.45 T$^{-1}$ in the alloy).
As seen in the figure, in both cases the most prominent peak among the Nernst anomalies
occurs at the quantum limit (9T in bismuth and 3T in the alloy). Note also the broadening
of the Nernst anomalies in the alloy. As for the ultraquantum regime, as seen in the figure,
three sharp anomalies were resolved in bismuth\cite{behnia2}. They occurred at fields corresponding
to $3/2$, $5/2$ and $7/2$ times the quantum limit. Here, in Bi$_{0.96}$Sb$_{0.04}$, we resolve
at least one ultraquantum anomaly at twice the quantum limit. We recall that
this field scale has very visible signatures in charge transport (See Fig. 2).

Can this anomaly be caused by the electron pockets? Since the latter
remain undetected, this question cannot be answered with certainty.
However, as seen in Fig. 2 the sign of the jump at 6 T in the Hall
resistivity is the same as the 3T anomaly: It is \emph{positive}. A
similar feature was seen in the Seebeck data. Both these suggest
that this anomaly is linked to the hole-like carriers. In the case
of pure bismuth, the electron pockets, which are elusive in
transport measurements, dominate the magnetization response and
torque magnetometry has emerged as a powerful probe of the electron
pockets across the quantum limit\cite{luli,fauque}. Future angular
torque magnetometry studies should shed more light on the origin of
the ultraquantum anomalies detected here.

In summary, we studied the transport properties of
Bi$_{0.04}$Sb$_{0.96}$ across the quantum limit and quantified the
carrier density and  mobility of this semimetal. Even at such low
level of carrier concentration, the zero-field ground state is a
Fermi liquid. In the extreme quantum limit, at least one anomaly
possibly emanating from unknown many-body effects was detected. We
thank G. Lapertot for precious technical assistance. This work was
supported by the Agence Nationale de la Recherche.

\end{document}